\documentclass[
    ,final            % use final for the camera ready runs
  ]
  {aipproc}

\layoutstyle{6x9}

\newcommand{\NP}[1]{{\it Nucl.\ Phys.\ } {\bf #1}}
\newcommand{\ZP}[1]{{\it Z.\ Phys.\ } {\bf #1}}

\newcommand{\PL}[1]{{\it Phys.\ Lett.\ } {\bf #1}}

\newcommand{\PR}[1]{{\it Phys.\ Rev.\ } {\bf #1}}
\newcommand{\PRL}[1]{{\it Phys.\ Rev.\ Lett.\ } {\bf #1}}

\newcommand{\Od}{{\cal O}}
\newcommand{\im}{\mbox{Im}\,}
\newcommand{\re}{\mbox{Re}\,}
\newcommand{\ov}{\overline}
\newcommand{\lsim}{\raise.3ex\hbox{$<$\kern-.75em\lower1ex\hbox{$\sim$}}}

\begin{document}

\begin{flushright}
LBNL-51770
\end{flushright}

\title{Thermal Meson properties within Chiral Perturbation Theory}

\author{A.G\'omez Nicola}{
  address={Departamento de F\'{\i}sica Te\'orica II,
  Universidad Complutense, 28040 Madrid, Spain.}
}

\author{J.R.Pel\'aez}{address={Dip. di Fisica. Universita' degli Studi and INFN,
 Firenze,
  Italy.},
  altaddress={Departamento de F\'{\i}sica Te\'orica II,
  Universidad Complutense, 28040 Madrid, Spain.}}

\author{A.Dobado}{
  address={Departamento de F\'{\i}sica Te\'orica I,
  Universidad Complutense, 28040 Madrid, Spain.}
 ,altaddress={Theory Group, Lawrence Berkeley National
 Laboratory,
Berkeley, CA 94720, USA.} % additional visiting address
}

\author{F.J.Llanes-Estrada}
{
  address={Departamento de F\'{\i}sica Te\'orica I,
  Universidad Complutense, 28040 Madrid, Spain.}
 % ,altaddress={<author1 address>} % additional visiting address
}

\begin{abstract}
 We report on our recent work about the description of a meson
  gas below the chiral phase transition within the framework
   of Chiral Perturbation Theory. As an alternative to the standard
   treatment, we present a calculation of the quark condensate
   which combines the virial expansion and the meson-meson scattering data.
    We have also calculated the
  full one-loop elastic pion scattering amplitude at finite
  temperature  and we have unitarized the amplitude using the
  Inverse Amplitude Method in order to reproduce the temperature effects on
   the mass and width of the $\rho$ and $\sigma$
   resonances.
  Our results show a clear increase of the thermal $\rho$ width, as expected
   from previous analysis. The results for the $\sigma$ are consistent
    with Chiral Symmetry Restoration.  We comment on  the relevance of our results
  within the context of Relativistic Heavy Ion Collisions.
 \end{abstract}

\maketitle

\section{Introduction and Motivation}

The recent development of the Relativistic Heavy Ion Collision
(RHIC) program is one of the main motivations to study hadronic
matter under extreme conditions of temperature $T$ and density.
Here we will consider the meson gas formed when the plasma created
after one such RHIC has expanded and hadronized, being in a state
where the chiral symmetry is restored. There are strong
indications that one could observe the medium effects on such a
system. For instance, the dilepton spectrum shows an anomalous
behaviour for invariant masses  near the $\rho$ mass
\cite{helios95}. The flatness of the spectrum is consistent with a
modification of the mass and width of the $\rho$'s which  have
time to decay inside the plasma, so that their spectral function
acquires thermal corrections due to the interaction with the hot
and dense hadron gas
\cite{de90,dom93,pis95,haglin,li,soko96,ele01,NJL1}.

In such a system, external momenta and temperature  are small
compared to the chiral symmetry breaking scale
$\Lambda_\chi\simeq$ 1 GeV. The relevant degrees of freedom are
then the lightest mesons and the interactions among them are best
described by a low-energy QCD effective theory based on chiral
symmetry. The most general framework comprising the  QCD Chiral
Symmetry Breaking pattern $SU_L(N_f)\times SU_R(N_f)\rightarrow
SU_V(N_f)$ is Chiral Perturbation Theory (ChPT)
\cite{we79,gale84,gale85,chptrev} where any observable can be
calculated as an expansion in $p/\Lambda_\chi$,  $p$ denoting a
meson mass, momenta or the temperature. Thus, ChPT provides
model-independent predictions, just by fixing a few low-energy
constants (LEC). This program has included the calculation of
thermodynamic observables such as the free energy density and the
quark condensate, as we will briefly review below. Throughout this
paper, we will neglect finite baryon density effects, as ideally
in the central rapidity region formed after a RHIC.

Since ChPT is intended to provide a systematic low-energy and
low-$T$ perturbative expansion, we do not expect it to reproduce
resonances. This is strongly related to the fact that ChPT
 only satisfies unitarity in a perturbative fashion.
 Over the last few years, there
has been a lot of work devoted to enlarge the ChPT applicability
range
 by using unitarization methods, i.e, imposing exact unitarity
 requirements, like the Inverse Amplitude Method (IAM)
 \cite{iam} or approaches based on Lippmann-Schwinger or
 Bethe-Salpeter equations \cite{BS}. These methods provide a good
 agreement with the experimental phase shifts and they are able to
 generate resonances, like the $\rho$ and the $\sigma$ for the
 $SU(2)$ chiral symmetry. In addition, they can be extended to
 include coupled channels, describing
  successfully all the meson-meson data and resonances
  in the $SU(3)$ case, up to
 1.2 GeV \cite{CC,gope02}.

What we will show below is that only requiring chiral symmetry and
unitarity one can also describe successfully the thermal behaviour
of the $\rho$ and $\sigma$ resonances. For that purpose, one needs
first to calculate the thermal pion scattering in ChPT, which has
been done in \cite{pipiT}. We shall discuss the main features of
such thermal amplitude below. Then, by using the IAM extended to
finite $T$, one can construct a nonperturbative thermal unitarized
amplitude  which, in particular, describes the behaviour of the
thermal $\rho$ and $\sigma$ \cite{rhoT}. We will present the
results for the thermal mass and width of the $\rho$ and $\sigma$
in that approach, as well as for the $T$-dependence of the
effective $g_{\rho\pi\pi}$ vertex. The main implications of our
results in the context of RHIC and  chiral symmetry restoration
will be also discussed below.

\section{The meson gas and Chiral Symmetry at finite $T$ within ChPT}

For the reasons explained above, it is important to provide an
accurate description of  the low-$T$ meson gas in thermal
equilibrium. For instance, the signature of chiral symmetry
restoration at $T_c\simeq$ 150-200 MeV should be observed in the
thermal evolution of the order parameter, the quark condensate
$\langle \bar q q \rangle (T)$ from below the transition point. As
we have just  discussed, ChPT provides a model independent
description of the thermodynamic observables, based only on chiral
symmetry. The only extra ingredient is the temperature, which is
treated as an $\Od(p)$ parameter.

The first calculations of the pion gas within ChPT go back to
\cite{gale87}, where $\langle \bar q q \rangle (T)$ and the
thermal dependence of $f_\pi (T)$ were calculated up to $\Od
(T^2)$ (one loop) in the chiral limit. That  result already showed
a behaviour compatible with chiral symmetry restoration. In
\cite{gele89} a thorough analysis up to $\Od (T^6)$ was performed,
including the free energy, the quark condensate and an estimate of
the thermal effects of {\it free} kaons and etas. The $\Od (T^4)$
corrections to $f_\pi (T)$ have been analyzed in \cite{fpifourth}
where it has been shown that beyond $\Od (T^2)$ one has to
consider separately the space and time components of the axial
current, so that there are two independent $f_\pi^{s,t}$, which,
in addition, can be complex. In fact, their imaginary part is
proportional to the in-medium pion damping rate while their real
parts are related to the  deviations of the  pion velocity from
the speed of light. Other analysis of the thermal pion dispersion
law can be found in \cite{gole89,schenk93}. The analysis of
typical nonequilibrium effects such as explosive pion production
after a RHIC can be also studied within the ChPT context
\cite{neq}. It should be pointed out that many of these properties
have also been investigated using specific models for low-energy
QCD. The most popular is
 the $O(4)$ model, which reproduces a critical behaviour already
 in the mean field approach. Apart from introducing the $\sigma$
 explicitly, conventional perturbation theory in the
$O(4)$ model breaks down, which has been dealt with at finite $T$
using different nonperturbative approaches \cite{onmod,chihat}.

When calculating thermodynamic quantities such as the pressure or
the quark condensate  from ChPT, the usual approach  is to use the
Feynman rules of Thermal Field Theory \cite{lebellac} to the order
considered. This is particularly cumbersome in the three flavor
case  if one wishes to include the full dependence on temperature,
quark masses and $SU(3)$ interactions. An alternative
\cite{gele89,dope99,pe02} is to perform a virial expansion of the
pressure as \cite{dashen}

\begin{equation}
\beta P=\sum_i B_{i}(T)\xi_i + \sum_i\left( B_{ii} (T) \xi_i^2 +
\frac{1}{2}\sum_{j\neq i} B_{ij} (T) \xi_i\xi_j \right)...,
\label{virialexp}
\end{equation}
where $i=\pi,K,\eta$. This is a dilute gas expansion in the
fugacities $\xi_i=\exp(-m_i/T)$. The binary interactions between
the different species show up in \cite{gele89,dope99,pe02}:

\begin{equation}
B_{ij}^{(int)}=\frac{\xi_i^{-1}\xi_j^{-1}}{2\,\pi^3}\int_{m_i+m_j}^{\infty}
dE\, E^2 K_1(E/T) \sum_{I,J,S} (2I+1)(2J+1)\delta^{ij}_{I,J,S}(E),
\label{2vircoef}
\end{equation}
where $K_1$ is the first modified Bessel function and
$\delta^{ij}_{I,J,S}$ are the elastic scattering $ij\rightarrow
ij$ phase shifts {\em at T=0} (chosen so that $\delta=0$ at
threshold) of
 a state $ij$ with  isospin $I$,  angular momentum $J$ and
 strangeness $S$. What makes the virial expansion useful is that
  the $T$ dependence on thermodynamical observables up to $T\simeq
 200-250$ MeV can be obtained just from the $T=0$ phase shifts, which have been
 calculated for all possible meson-meson interactions in $SU(3)$
 ChPT to one loop. They can be found for instance in
 \cite{gope02}. Note that for the pressure one could even use the
 experimental phase shifts directly. However, the quark condensate is
  given by the derivative of the pressure with respect to the quark
  mass and therefore  the analytic dependence of the
  $\delta^{ij}$ with the different meson masses is needed.

For temperatures much below 150 MeV, massive states like kaons and
eta are thermally suppressed, typically  by the Boltzmann  factors
$\exp(-M_K/T)$ and, in principle, the suppression is even stronger
for the interactions among them and with pions, as
(\ref{virialexp}) shows. For low $T$ it is then reasonable to
treat those states as free particles, as it was done in
\cite{gele89}. Moreover, at low $T$ the integrals in
(\ref{2vircoef}) are dominated by the phase shifts at threshold.
However, for higher temperatures, the effect of massive states
becomes increasingly important and, furthermore, the dependence of
the interactions with the pion mass can be large
 so that their contribution to the quark condensate
becomes sizable \cite{pe02}. When the effect of the strange states
is taken into account, there are  two main questions that can be
analyzed: on the one hand, the effect  on the non-strange
condensate $\langle\bar u u\rangle=\langle\bar d
d\rangle=\langle\bar q q \rangle/2$ (in the isospin limit) of
adding another flavor. Since the number of degrees of freedom
increases, so it does the entropy and one expects that the
collective state is closer to "disorder". This would imply a
decrease of the critical temperature, as it is indeed observed in
lattice calculations \cite{karsch}, although with gluons and
 massless quarks. On the other hand, one can
study the thermal evolution of the strange condensate $\langle\bar
s s\rangle$. Since $m_s\gg m_{u,d}$, one expects that the thermal
evolution for $\langle\bar s s\rangle (T)$ is slower than for the
non-strange condensate. Note that the quark masses play a similar
role as external magnetic fields in ferromagnets, so that
increasing the external field means that it takes more thermal
energy to restore the symmetry. Let us remark that, in contrast to
the lattice, the physical masses are easily incorporated in our
approach.

\begin{figure}
\includegraphics[height=.3\textheight]{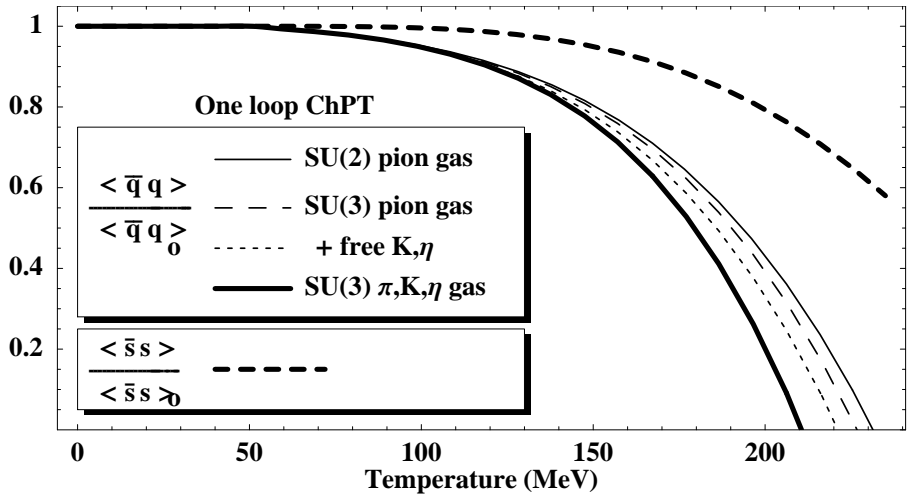}
\caption{The thermal evolution of the $\langle\bar q q \rangle$
and $\langle\bar s s \rangle$ condensates in ChPT}
\label{fig:cond}
\end{figure}

We report here on the virial expansion calculation in $SU(3)$ ChPT
that has been done in \cite{pe02}. The main results for the quark
condensates are summarized in Figure \ref{fig:cond}. First, it
should be pointed out that the curves are plotted at most up to
the point where  the condensates vanish. The ChPT  condensates do
not vanish above those "critical" values, since they have been
obtained from a perturbative series in the temperature. In fact,
the results should be trusted only in the region showed in the
graphs, as explained above. The thermal evolution of $\langle\bar
q q \rangle (T)$ is shown in different approximations, namely,
considering only pions in $SU(2)$ or $SU(3)$ (their tiny
difference comes from the $\Od(p^4)$ phase shifts), adding free
kaons and etas \cite{gele89} and, finally, considering the full
$SU(3)$ interactions \cite{pe02}. Note also that these
interactions (basically only $\pi K$ and $\pi\eta$ are important
at these temperatures) yield a larger effect than naively
expected. The reason is that they depend strongly on $m_\pi$,
which is more sensitive to $m_{u,d}$ than $m_K$ or $m_\eta$.
Taking into account the numerical values of the LEC's with their
errors, one gets a reduction of the melting temperature of
$T_m^{\langle \bar{q} q\rangle SU(2)}-T_m^{\langle \bar{q}
q\rangle SU(3)} = 21^{+14}_{-7}$ MeV, in remarkable agreement with
the chiral limit lattice calculations \cite{karsch}, taking into
account that we have used the actual meson masses. In addition, we
estimate $T_m^{\langle \bar{q} q\rangle SU(3)} = 211^{+19}_{- 7}$
MeV \cite{pe02}. The second effect, is also clearly seen in Figure
\ref{fig:cond}: The thermal evolution of the strange condensate
from the broken phase is much slower than the non-strange one.
From Figure \ref{fig:cond}, we see that there is still about 80 \%
left for $\langle \bar s s\rangle (T)/\langle \bar s s\rangle (0)$
when $\langle \bar q q\rangle (T)/\langle \bar q q\rangle (0)$ has
already melted. Finally, one may wonder about the effect of
calculating the integrals in (\ref{2vircoef}) with ChPT, whose
applicability does not extend to infinity. Indeed, when
(\ref{2vircoef}) is evaluated \cite{pe02} with the phase shifts
unitarized with the coupled channel IAM \cite{gope02} , which have
a much larger applicability range, the numerical results only
change very slightly. Thus, the main conclusions remain the same,
since, as we have already commented, the main contribution to the
integrals comes from the low-energy region and the IAM agrees with
ChPT at low energy, improving only the high energy behavior.

\section{Pion scattering at $T\neq 0$ in one loop ChPT}

 If the pion gas is dilute enough, i.e., at very low temperatures,
  it is reasonable to assume that
 the only dependence of pion scattering with $T$ can be accounted for
  through the Bose-Einstein distribution functions and one can
  ignore the $T$-dependence of the scattering amplitudes. However,
   that dependence could be important in several contexts. For
   instance, it has been suggested that an enhancement of pion scattering in the scalar
   channel near threshold could be an indication of chiral symmetry restoration
 \cite{chihat}. In addition, a previous calculation \cite{NJL1,NJL} of thermal $\pi\pi$
 scattering in the Nambu-Jona-Lasinio model,
  shows a singular behaviour of the
 scattering lengths at some critical temperature, which may be
 related to a Mott transition.
Furthermore, if one
 wishes to extend to finite $T$ the fruitful unitarization program in ChPT to
 describe resonances, the full calculation of the $\pi\pi$
 scattering amplitude to one loop and the extension of perturbative unitarity are
 essential ingredients.

  For the above reasons, it is important to provide a
 model-independent description of pion scattering for temperatures
 $T$ well below the chiral scale $\Lambda_\chi$. This can be
 achieved in ChPT.  The calculation of the thermal scattering
 lengths in ChPT to one loop has been done in \cite{Kaiser}
 whereas the calculation of the full thermal amplitude in one loop
 ChPT has been recently carried out \cite{pipiT}. We will report
 here the main results and  features of that work.

 There are two formal aspects regarding the calculation of the
 scattering amplitude at finite $T$ that are worth pointing out.
 The first one is that we are considering the thermal amplitude
 defined by taking $T=0$ asymptotic pion sates,
 the $T$ dependence coming from the four-point function,
 calculated using the Feynman rules of Thermal Field Theory in the
 imaginary-time formalism \cite{lebellac}. Then, one can perform
 an analytic continuation from discrete frequencies
 $\omega_n=2\pi n T$ to  real energies $E$ as
 $i\omega_n\rightarrow E+i\epsilon$. Such analytic continuation
 corresponds to the retarded four-point function in the so called
 real-time formalism \cite{bani94}. The retarded functions have a
 causal and analytic structure \cite{kobes} suitable  to extend
 perturbative unitarity at $T\neq 0$. The same definition of
 thermal amplitude has been used in \cite{NJL1,Kaiser}.

\begin{figure}
\includegraphics[height=.3\textheight]{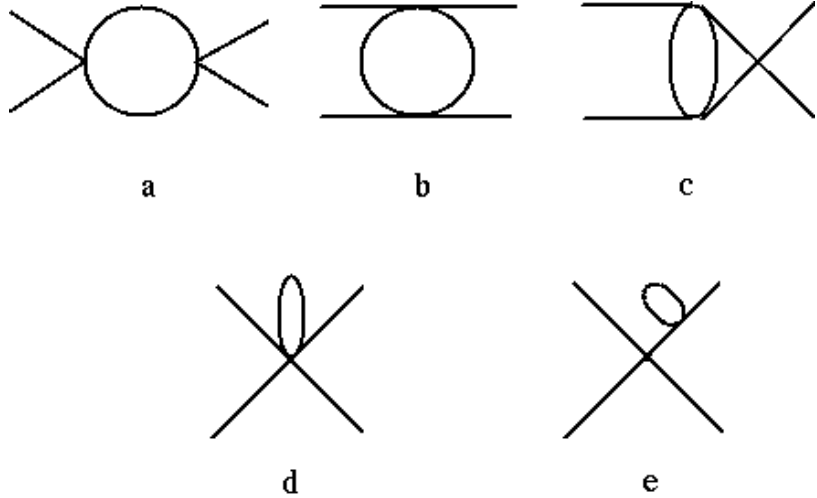}
\caption{One loop diagrams contributing to $\pi\pi$ scattering}
\label{fig:diag}
\end{figure}

The second point is that the loss of Lorentz covariance inherent
to the thermal formalism (due to the choice of the thermal bath
rest frame) means that any two-body scattering amplitude with
four-momenta $k_1  k_2\rightarrow k_3 k_4$ will depend separately
on the variables $ \bf{S_0}$, $\vert \vec{ \bf{S}}\vert$, $
\bf{T_0}$, $\vert \vec{ {\bf T}}\vert$, $ \bf{U_0}$ and $\vert
\vec{ {\bf U}}\vert$, where ${\bf S}=k_1+k_2$, $ {\bf T}=k_1-k_3$
and $ {\bf U}=k_1-k_4$. At $T=0$, the amplitude depends only on
the Mandelstam variables $s={\bf S^2},t={\bf T^2},u={\bf U^2}$ and
any $\pi\pi$ scattering amplitude can be related to that of
$\pi^+\pi^-\rightarrow\pi^0\pi^0$, called  $A(s,t,u)$, by isospin
and crossing transformations. At $T\neq 0$ and since the
temperature does not modify the interaction vertices, the crossing
symmetry still holds, but now  in terms of ${\bf S}$, ${\bf T}$
and ${\bf U}$. Therefore,   any $\pi\pi$ thermal amplitude can be
written in terms of the thermal $A({\bf S},{\bf T},{\bf
U};\beta)$. The loop  diagrams are the same as for $T=0$ and are
given in Figure \ref{fig:diag}.

The final result for the thermal amplitude can be written as:

 \begin{equation} A({\bf S},{\bf T},{\bf U};\beta)=A_2
({\bf S},{\bf T},{\bf U}) + A_4^{pol}({\bf S},{\bf T},{\bf U})
\label{Aparts}+ A_4^{tad} ({\bf S},{\bf T},{\bf
U};\beta)+A_4^{uni} ({\bf S},{\bf T},{\bf U}; \beta),
\end{equation}
where   $A_2$ is  tree level contribution coming from the
$\Od(p^2)$ lagrangian (the nonlinear sigma model) while
$A_4^{pol}$ contains both the tree level $\Od(p^4)$ contributions
plus polynomials coming from the renormalization of the loop
integrals. Both $A_2$ and $A_4^{pol}$ are temperature independent.
The $A_4^{uni}$ term represents those contributions from diagrams
a,b,c in Figure \ref{fig:diag}. They  yield the correct analytic
structure and will ensure perturbative unitarity in all channels.
Finally, the contribution $A_4^{tad}$ accounts for  tadpoles like
those in diagrams d,e in Figure \ref{fig:diag} plus
 terms coming from diagrams a,b,c  proportional to
 the tadpole integral.

 The detailed
 results for the different contributions above can be found in
 \cite{pipiT} and we do not give them here for brevity. As a first check
 of consistency, we recover the $T\rightarrow 0^+$ limit of
 \cite{gale84}. Furthermore, when the thermal amplitude is
 projected into partial waves $a_{IJ}$ of definite isospin $I$ and
 angular momentum $J$ (defined in the center of mass frame where the pions are at
  rest with the thermal bath) we also agree with the results given in
 \cite{Kaiser} for the scattering lengths.

 Another important check
 of consistency, which will be crucial for our analysis in the next section,
 concerns the imaginary part of the partial waves
 and perturbative unitarity. At $T=0$, unitarity constraints the partial waves, for $s>4m_\pi^2$
and below other inelastic thresholds, to satisfy
\begin{equation}
 \im a(s)=\sigma (s) \vert a (s) \vert^2, \label{unit0}
 \end{equation} where  $\sigma (s)=\sqrt{1-4 m_\pi^2/s}$
 is the two-pion phase space,
whereas the ChPT series only satisfies unitarity perturbatively,
i.e.,
\begin{equation}
 \im a_2 (s)= 0, \quad \im a_4 (s) = \sigma (s)
\left\vert a_2(s) \right\vert^2, \quad  .... \label{pertunit0}
\end{equation}

\begin{figure}
\includegraphics[height=.5\textheight, angle=-90]{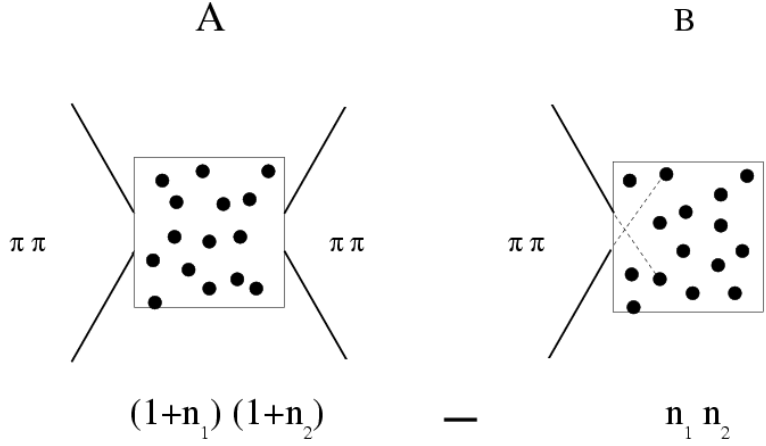}
\caption{Interpretation of thermal bath contributions to the pion
scattering phase space. In the process A, the medium enhances the
two-pion production, while process B represents the absorption of
the initial pions by the thermal bath.} \label{fig:bath}
\end{figure}

It is possible to generalize the relation (\ref{pertunit0}) to the
 case of any one-loop elastic scattering amplitude at finite $T$.
 The derivation is given in \cite{pipiT} and follows closely the analysis
 in \cite{weldon} of the discontinuity in the self-energy
 of a particle decaying in the thermal bath. The result for the
 thermal perturbative unitarity relation is:

\begin{equation}
\label{pertunitT} \im a_2 (s)= 0, \qquad \im a_4 (s;\beta) =
\sigma_T ({\bf S_0}) \left\vert a_2(s)\right\vert^2\ , \quad {\bf
S_0}>2m_\pi,
\end{equation}
where
\begin{equation}
\sigma_T (E)=\sigma(E^2)\left[1+\frac{2}{\exp({\beta\vert E
\vert/2})-1}\right] \label{sigmaT}
\end{equation}
and it has been assumed that only $\pi\pi$ states are available in
the thermal bath. Thus, $\sigma_T$ is nothing but the thermal
phase space, whose origin can be physically interpreted by writing
$1+2n_B(E/2)=\left[1+n_B(E/2)\right]^2-n_B^2(E/2)$ where
$n_B(x)=\left(\exp(x/T)-1)\right)^{-1}$ is the Bose-Einstein
distribution function. Written in this way, the first term
represents the enhancement of phase space due to the presence of
pions in the medium, while the second term accounts for the
absorption of the two initial pions by pions in the bath. This is
schematically depicted in Figure \ref{fig:bath}. We have checked
explicitly that the relation (\ref{pertunitT}) holds for our
thermal $\pi\pi$ scattering amplitude. Moreover, the partial waves
at $T\neq 0$ can be  analytically continued  to the $s$ complex
plane \cite{rhoT} and they display the same analytic structure as
the $T=0$ one, i.e, they have a right unitarity cut in the real
axis above $s>4m_\pi^2$ (the discontinuity across the cut is given
by (\ref{pertunitT})) and a left one for $s<0$.
 We remark that our results remain valid when the density of
 states with more than two pions is small. This is equivalent to
 neglecting  higher powers of $n_B$ and is therefore similar to the
 dilute gas approach. For small energies and temperatures, that
 is, strictly within ChPT, the dilute gas approach is consistent,
 as we have also seen in the previous section. When the range of
 energies is extended, as we will do in the next section, one
 should bear in mind that the range of $T$ where our approach is
 valid is such that the Bose-Einstein factors still remain small.

\begin{figure}
\includegraphics[height=.22\textheight]{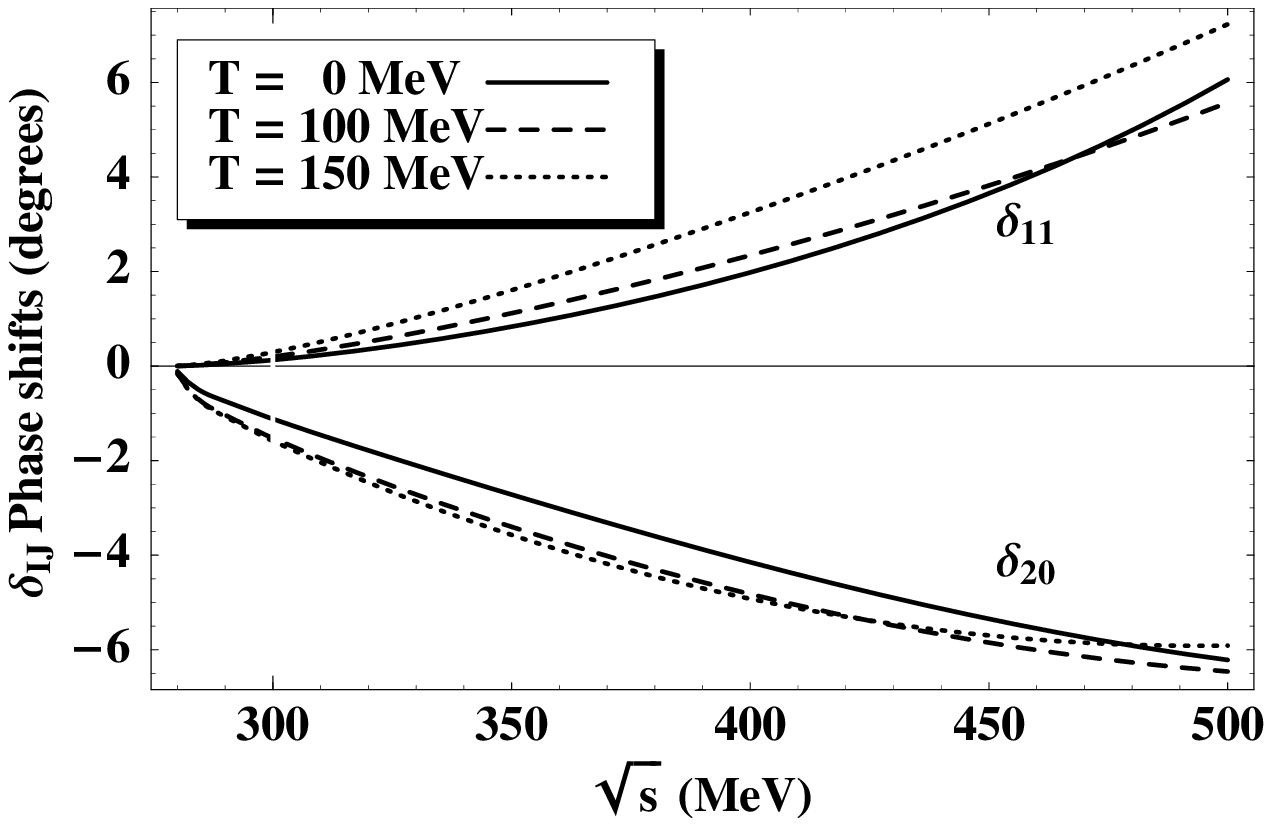}
\includegraphics[height=.22\textheight]{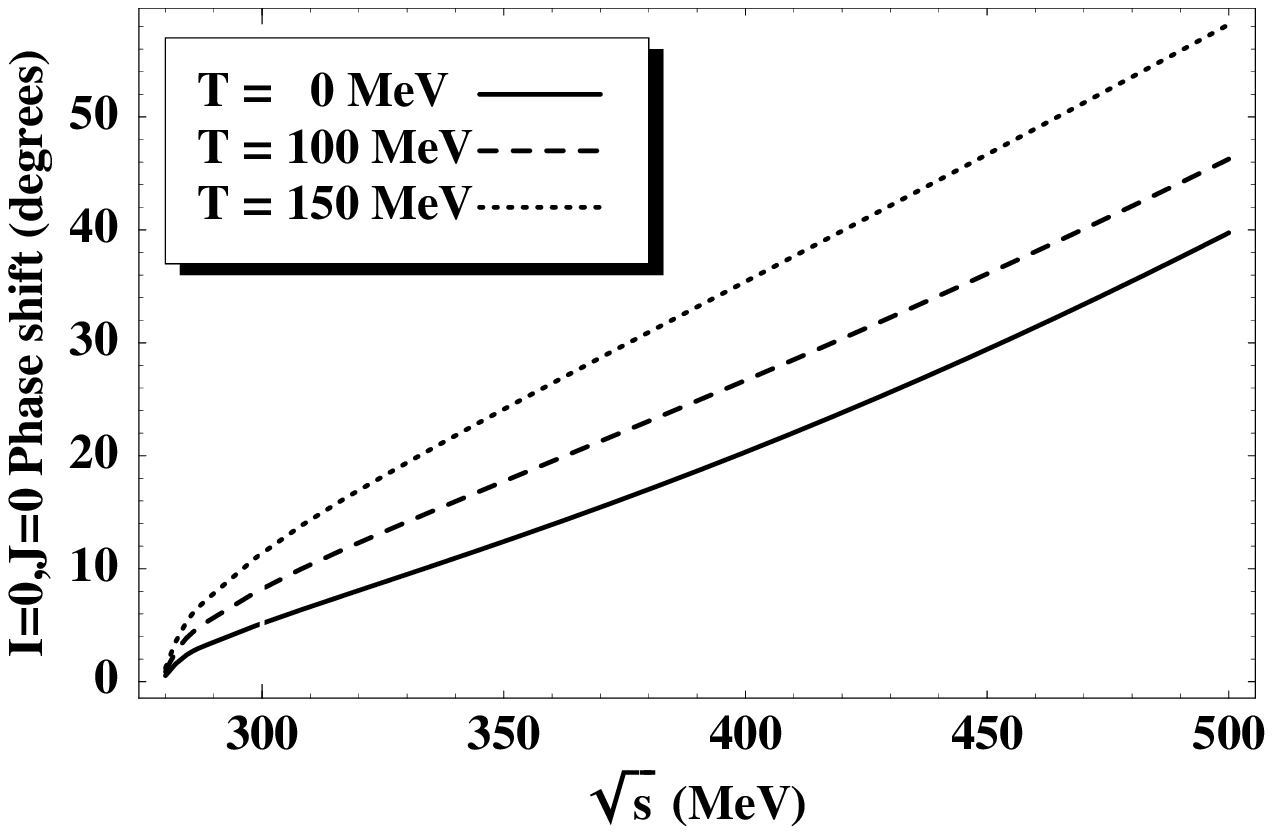}
\caption{Temperature evolution of the phase shifts $\delta_{IJ}$
for  $IJ=11,00,20$.} \label{fig:Tps}
\end{figure}

Finally, we have plotted the thermal phase shifts in Figure
\ref{fig:Tps}. These results deserve some comments. First, we
observe that the absolute value of the phase shifts in all
channels increases with $T$, while their sign (i.e, the attractive
or repulsive nature of each channel) is preserved. Recall that the
phase shifts are related to the real part of the amplitude as
$\delta_T\simeq \sigma_T \left(\sqrt{s}\right)\left[a_2(s)+\re
a_4(s,\beta)\right]$. The dominant contribution to the phase
shifts thermal enhancement is given by the thermal phase space
factor $\sigma_T$, while the $T$-dependence of the real part of
the amplitude is rather weak at low $T$. In particular, we do not
observe any significant thermal enhancement for the real part of
$a_{00}$, which would be interpreted as chiral symmetry
restoration \cite{chihat}. However, it must be stressed that we
are just considering the {\em perturbative} amplitude, valid only
for low $T$. In the next section, we will consider a
nonperturbative extension of the amplitude, which does show a
behaviour compatible with chiral symmetry restoration in the $00$
channel.

Another important comment regarding our results shown in Figure
\ref{fig:Tps} concerns the 11 channel, i.e, the $\rho$ channel.
Recall that, following the hypothesis of resonance saturation,
increasing $\delta_{11}$ is equivalent to increasing  the ratio
$\Gamma_\rho f_\pi^4/M_\rho^5$ \cite{gale84,Res}. Therefore, our
results are consistent with a  thermal increase of the rho width
coming mostly from thermal phase space and an almost constant rho
mass, at very low $T$ \cite{dom93,pis95,ele01}. This behaviour
will be confirmed by our analysis in the next section, where we
will find also important corrections at higher temperatures.

\section{The thermal $\rho$ and $\sigma$}

The purpose of this section is to show that only from chiral
symmetry and unitarity one finds a  thermal evolution of the
masses and widths of the $\rho$ and $\sigma$ at rest consistent
with previous analysis and with the dilepton spectrum observed in
RHIC.
 Unlike the models where the resonances are included as explicit
 degrees of freedom, we will start from the model-independent pion
 scattering amplitude in one loop ChPT calculated in the previous
 section and, imposing exact unitarity, we will construct a
 nonperturbative amplitude whose poles in the second Riemann sheet
  in the $00$ and $11$ channels correspond to the $\sigma$ and
  $\rho$ respectively.

  The exact unitarity relation (\ref{unit0}) implies that any partial wave
  satisfies $a=1/(\re a^{-1}-i \sigma)$ on the real axis below inelastic
thresholds. A unitarization method is just one way of
approximating $\re a^{-1}$. The IAM uses the one-loop ChPT and
thus ensures  that  exact unitarity is exactly satisfied and, at
the same time, the low-energy predictions of ChPT are preserved.
The IAM unitarized amplitude for one channel reads
$a^{IAM}=a_2^2/(a_2-a_4)$ \cite{iam} and coincides formally with
the [1,1] Pad\'e approximant in the $f_\pi^{-2}$ expansion.

We have already sketched in the introduction the virtues of the
IAM formula at $T=0$ for the description of resonances and the
data for higher energies. At $T\neq 0$, we have seen in the
previous section that, perturbatively in ChPT and for a dilute
gas, the partial waves satisfy thermal unitarity
(\ref{pertunitT})-(\ref{sigmaT}). Therefore, following the same
steps as for the $T=0$ case and motivated by the success of the
IAM approach, we will consider the unitarized IAM thermal
amplitude:

\begin{equation} \label{theriam}
a^{IAM}(s;T)=\frac{a_2^2(s)}{a_2(s)-a_4(s;T)},
\end{equation}  which satisfies the exact elastic unitarity
condition
\begin{equation} \label{iamunittf}
\im a^{IAM} (s;T)=\sigma_T (s) \left\vert
a^{IAM}(s;T)\right\vert^2.
\end{equation}

The first hint that (\ref{theriam}) provides a proper description
of thermal resonances comes from the following simple argument.
The behaviour of the $11$ partial wave in the real axis near
$s=M_T^2$, the $\rho$ mass squared,  can be described by a
Breit-Wigner parametrization (valid for narrow resonances
$\Gamma_T\ll M_T$):

\begin{equation} \label{brewig}
a^{BW}(s;T)=\frac{R_T(s)}{s-M_T^2+i\Gamma_TM_T}
\end{equation}
where $R_T(s)$  can be related to the $\rho\pi\pi$ effective
vertex (see below). Now,  compare  (\ref{theriam}) with
(\ref{brewig}) near $s=M_T^2$.  First, one gets $\re a_4
(M_T^2)=a_2(M_T^2)$, which defines the resonance mass and, second,
from the unitarity relation (\ref{iamunittf}), we have $\Gamma_T
M_T=-R_T (M_T^2) \sigma_T (M_T^2)$. Therefore,  if the thermal
corrections to $R_T$ and to $M_T$ were much smaller than those to
$\Gamma_T$ ($R_T \simeq R_0$ and $M_T\simeq M_0$) we would
simply get

\begin{equation} \label{gammaphase}
\Gamma_T\simeq \Gamma_0 \left[1+2 n_B\left(M_0/2\right)\right],
\end{equation}
which is the behaviour expected at very low $T$ for a $\rho$ at
rest \cite{pis95,dom93,ele01} that we had already anticipated in
the previous section. As discussed in \cite{weldon,pis95} and also
in the previous section, this result accounts for the difference
between the direct decay $\rho\rightarrow \pi\pi$ and the inverse
one $\pi\pi\rightarrow\rho$ which is allowed in the thermal bath
and is responsible for the dilepton production. Note that in order
to derive (\ref{gammaphase}) we have neglected the $T$-dependence
in $\re a_4 (S;T)$. Therefore, by using the complete result for
the thermal one-loop amplitude discussed in the previous section,
we will be able to find, for higher temperatures, the $T$
dependence on $M_T$, $R_T$, possible deviations from the low-$T$
behaviour (\ref{gammaphase}), and even more importantly, the
thermal evolution of both the $\rho$ and $\sigma$ poles in the
complex plane, which is the consistent way to generate resonances
within ChPT. Recall that a Breit-Wigner description is not valid
for the $\sigma$. According to our previous discussion, the upper
limit in $T$ to which our approach is valid will be dictated  by
the condition $n_B(M/2,T)<1$ where $M$ is the mass of the
resonance described ($\rho$ or $\sigma$). This gives roughly $T<$
300 MeV for the $\rho$ and $T<$ 180 MeV for the $\sigma$.

In the above discussion, it is crucial to obtain the analytic
continuation of the thermal amplitude to the complex $s$-plane.
The details can be found in \cite{rhoT}. The result shows the same
 right and left cuts structure as the $T=0$ amplitude, the discontinuities
 across the cuts being $T$-dependent.
 Once such analytic continuation has been  obtained, it can be
continued to the second Riemann sheet, using (\ref{iamunittf}), as
$a^{II}(s;T)=a^{IAM} (s;T)/[1-2i \sigma_T (s) a^{IAM} (s;T)]$.

\begin{figure}
\includegraphics[height=.3\textheight]{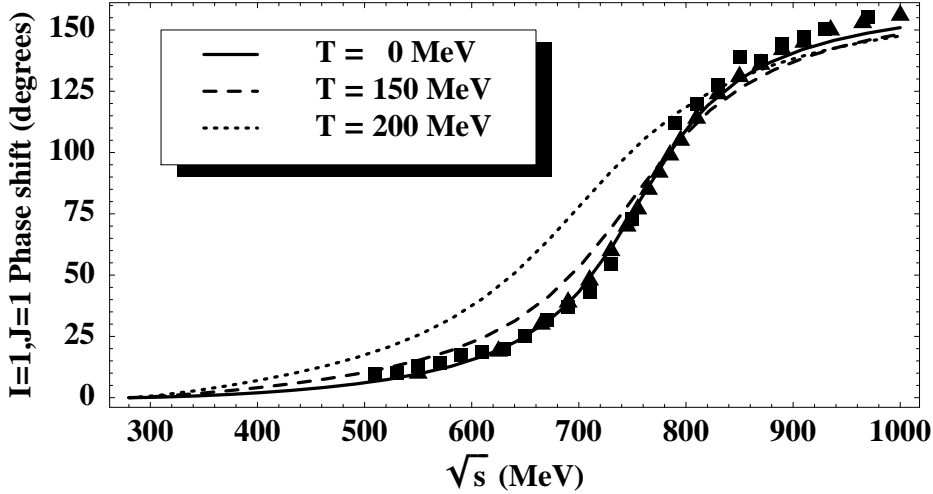}
\caption{$\delta_{11}$ for different temperatures. For the data
see \cite{iam}.} \label{fig:11uni}
\end{figure}

\begin{figure}
\includegraphics[height=.37\textheight]{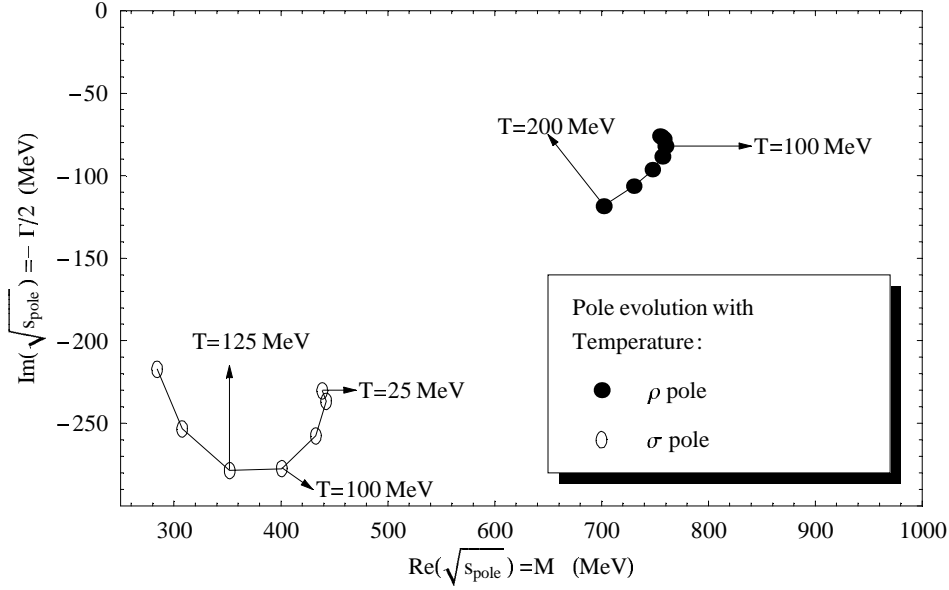}
\caption{Evolution of the $\sigma$ and $\rho$ poles with the
temperature.} \label{fig:poles}
\end{figure}

 Let us first show the results for $\delta_{11}$ for different
 temperatures, depicted in Figure \ref{fig:11uni}. The $SU(2)$ LEC
  (see \cite{gale84} for their definition)
 we have
 used are $\overline{l}_1=-0.3$, $\ov{l}_2=5.6$, $\ov{l}_3=3.4$ and
$\ov{l}_4=4.3$ and are obtained by fitting the $T=0$ scattering
data, which yields $M_0=$ 770 MeV and $\Gamma_0=$ 159 MeV. We see
clearly the broadening of the $\rho$ as $T$ increases. This is
confirmed by the evolution of the thermal poles, which is shown in
Figure \ref{fig:poles}. The $\rho$ width increases with $T$  while
the $\rho$ mass decreases slightly. The $\sigma$ pole deserves
some comments. We see that for low $T$ the $\sigma$ width
increases and its mass decreases slightly, following a similar
behaviour as the $\rho$, i.e, mostly dictated  by the thermal
space factor $\sigma_T$. However, for $T>$ 125 MeV,
$M_\sigma$
 decreases more rapidly and $\Gamma_\sigma$ starts to decrease. This behaviour
 has an interesting explanation in terms of chiral symmetry
 restoration. On the one hand, since the $T$-dependence of $m_\pi (T)$ is rather
 weak up to $T\simeq$ 200 MeV \cite{schenk93,chihat},
  the decrease of $M_\sigma (T)$ points towards
  $M_\sigma\rightarrow m_\pi$ and that decreasing is stronger as $T$
   increases, unlike the $\rho$ mass. On the other hand, as $T$ increases
  and $M_\sigma$ approaches $2m_\pi$ from above, both the direct
  $\sigma\rightarrow 2\pi$ and inverse $2\pi\rightarrow \sigma$
  decays become kinematically disallowed, so that
  $\Gamma_\sigma$ is reduced. At low $T$, the decrease of the mass
  is much weaker and the phase space contribution dominates,
  making $\Gamma_\sigma$ grow. Similar results and discussion
  for the $\sigma$ can be found
  in  \cite{chihat,yoko}.

\begin{figure}
\includegraphics[height=.3\textheight]{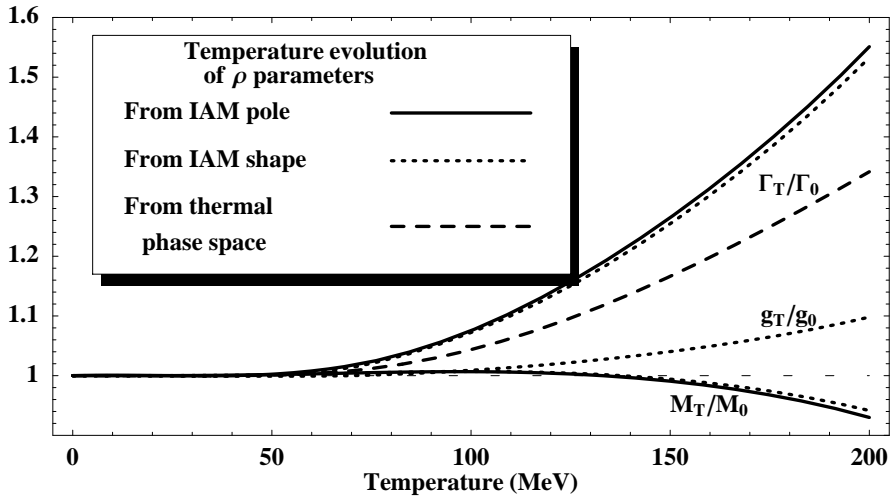}
\caption{Thermal evolution of the $\rho$ mass, width and
$\rho\pi\pi$ effective vertex} \label{fig:sum}
\end{figure}

Finally, in Figure \ref{fig:sum} we have collected our results for
the $\rho$. First, we clearly observe a significant deviation (for
$T>$ 100 MeV) between the full result (obtained either from the
pole position in Figure \ref{fig:poles} or from the phase shift in
Figure \ref{fig:11uni}) and the naive thermal phase space
prediction (\ref{gammaphase}), stressing the importance of having
a full $T$-dependent ChPT description of $\pi\pi$ scattering.
Moreover, although $M_\rho$ changes little, consistently with
Vector Meson Dominance \cite{de90,pis95}, our results show a
sizable slight decrease of the $\rho$ mass for $T>$ 150 MeV, which
seems to be favored by phenomenological analysis of RHIC dilepton
data \cite{li,ele01}. In Figure \ref{fig:sum} we have also plotted
the effective $\rho\pi\pi$ vertex, defined from $R_T$ in
(\ref{brewig}) as $R_T= g_T^2 \left(4m_\pi^2-M_T^2\right)/48\pi$,
 from
 the VMD $\rho\pi\pi$ coupling \cite{pis95,soko96}
  with a thermal $\rho$ ($g_0\simeq$ 6.2). At low temperatures
$g_T\lsim g_0$ ($g_{50}/g_0\simeq$ 0.9991) in agreement with the
chiral analysis in \cite{soko96}. For higher temperatures $g_T$
grows, although the thermal corrections are much smaller than
those at finite baryon density  \cite{broflohi01}.

\section{Conclusions}

We have reviewed some basic aspects about the ChPT description of
the meson gas formed after a Relativistic Heavy ion Collision. In
particular, we have shown that one can combine the virial
expansion with ChPT in order to obtain the thermal evolution of
the quark condensate below the chiral phase transition in the
dilute gas regime. That approach allows to use directly all the
available information on meson-meson scattering for three light
flavors in ChPT. Contrary to lattice studies, in this approach
physical masses are easily accounted for. The main effects of
considering the strange sector in the condensate are, first, that
the strange condensate melts much more slowly than the non-strange
one and, second, that the melting temperature for the non-strange
condensate is sizably smaller with three flavors than with two,
and that, surprisingly, the $\pi K$ and $\pi \eta$ interactions
provide a largest contribution to this effect than free kaons or
etas.

We have also shown the results of a recent calculation of the
$\pi\pi$ scattering amplitude at finite $T$ in one loop ChPT. The
partial waves satisfy an extended version of perturbative
unitarity, where the only change with respect to the $T=0$ case is
that the phase space is thermally enhanced by the presence of
pions in the thermal bath. Thus, the imaginary part of the thermal
amplitude has a neat physical interpretation in terms of
absorption and induced emission of pion pairs in the thermal bath.
The thermal phase shifts are also enhanced with $T$, keeping their
attractive or repulsive nature, mainly dominated by the phase
space factor.

The thermal pion scattering amplitude can be unitarized using the
IAM, following similar steps as for $T=0$. The unitarized
 partial waves can then be analytically continued to
the complex energy plane and their poles in the $I=J=1$ and
$I=J=0$ channels correspond to the $\rho$ and the $\sigma$. This
approach provides  a description of the thermal effects  for a
$\rho$ and $\sigma$ at rest, only from chiral symmetry and
unitarity. We have found that  $\Gamma_\rho$ increases
significantly with $T$. For low $T$, $M_\rho$ and the effective
vertex $g_{\rho\pi\pi}$ remain almost constant, which is
consistent with neglecting the $T$-dependence in the real part of
the amplitudes and considering only the contribution from the
thermal phase space. However, for higher $T$, the full
$T$-dependence in the amplitude has to be taken into account.
Thus,  for $T>$ 100 MeV, $\Gamma_\rho$ acquires significant
corrections from the pure thermal space approximation and
$g_{\rho\pi\pi}$  increases, while for $T$>150 MeV, $M_\rho$ shows
a sizable decreasing behaviour. As for the $\sigma$ pole, although
its interpretation as a resonance is much less clear than the
$\rho$, the thermal behaviour of both $M_\sigma$ and
$\Gamma_\sigma$, as obtained from its associated pole, can be
understood from chiral symmetry restoration.

Our results agree with several theoretical analysis and are
consistent with phenomenological studies of RHIC dilepton data. We
have only used chiral symmetry and unitarity as guiding
principles, without including the resonances as explicit degrees
of freedom. It would be interesting to apply our chiral unitary
approach to the production of dileptons from thermal $\pi\pi$
annihilation near the $\rho$ energy, in order to be able to
provide more accurate predictions about the observed dilepton
spectrum in RHIC. This is just but one of the possible directions
that we will pursue in the near future.

%%%%%%%%%%%%%%%%%%%%%%%%%%%%%%%%%%%%%%%%%%%%%%%%
%% BACKMATTER
%%%%%%%%%%%%%%%%%%%%%%%%%%%%%%%%%%%%%%%%%%%%%%%%

\begin{theacknowledgments}
A.G.N and J.R.P wish to thank the organizers of the "2nd
International  Workshop on Hadron Physics" for their kind
invitation. We acknowledge financial  support from  the Spanish
CICYT projects, FPA2000-0956, PB98-0782 and BFM2000-1326. This
work was supported in part by the Director, Office of Science,
Office of High Energy and Nuclear Physics of the U.S. Department
of Energy under Contract DE-AC03-76SF00098. J.R.P. acknowledges
support from the CICYT-INFN collaboration grant 003P 640.15. A.D.
acknowledges support from the Universidad Complutense del Amo
Program.
  \end{theacknowledgments}

%% If the bibliography is produced without BibTeX comment out the

\bibliographystyle{aipproc}   % if natbib is available

\end{document}